%% file: 0_main.tex
\documentclass[sigconf]{acmart}

\usepackage{enumitem}

\usepackage{amssymb}
\usepackage{pifont}
\usepackage{url}
\usepackage{multirow}
\usepackage{subfigure}
\usepackage{bm}
\usepackage{verbatim}
\usepackage{ragged2e}
\usepackage{caption}
\usepackage{subcaption}
\usepackage{graphicx}
\usepackage[normalem]{ulem}
\useunder{\uline}{\ul}{}
\usepackage{amsmath}
\usepackage{longtable}
\usepackage{adjustbox}
\usepackage{microtype}
\usepackage{setspace}
\usepackage{hyperref}
\usepackage{float}
\usepackage{balance}
\usepackage{algorithm}
\usepackage{algpseudocode}

\newcommand{\eg}{\emph{e.g., }}

\newcommand{\aka}

\newcommand{\bym}[1]{\textcolor{black}{#1}}

\AtBeginDocument{%
  }

\copyrightyear{2026}
\acmYear{2026}
\setcopyright{cc}
\setcctype{by}
\acmConference[SIGIR '26]{Proceedings of the 49th International ACM SIGIR Conference on Research and Development in Information Retrieval}{July 20--24, 2026}{Melbourne, VIC, Australia}
\acmBooktitle{Proceedings of the 49th International ACM SIGIR Conference on Research and Development in Information Retrieval (SIGIR '26), July 20--24, 2026, Melbourne, VIC, Australia}
\acmDOI{10.1145/3805712.3809632}

\acmISBN{979-8-4007-2599-9/2026/07}

\begin{document}

\title{Bi-Level Optimization for Generative Recommendation: Bridging Tokenization and Generation}

\author{Yimeng Bai}
\orcid{0009-0008-8874-9409}
\affiliation{
  \institution{University of Science and Technology of China}
  \city{Hefei}
  \country{China}
}
\email{baiyimeng@mail.ustc.edu.cn}

\author{Chang Liu}
\orcid{0009-0000-8324-7153}
\affiliation{
  \institution{Beihang University}
  \city{Beijing}
  \country{China}
}
\email{chang.liu@buaa.edu.cn}
\authornote{Equal Contribution.}

\author{Yang Zhang}
\orcid{0000-0002-7863-5183}
\affiliation{
  \institution{National University of Singapore}
  \city{Singapore}
  \country{Singapore}
}
\email{zyang1580@gmail.com}
\authornote{Corresponding author.}

\author{Dingxian Wang}
\orcid{0000-0002-6880-7869}
\affiliation{
  \institution{Upwork}
  \city{Seattle}
  \country{United States}
}
\email{dingxianwang@upwork.com}

\author{Frank Yang}
\orcid{0009-0001-8117-8788}
\affiliation{
  \institution{Upwork}
  \city{Seattle}
  \country{United States}
}
\email{frankyang@upwork.com}

\author{Andrew Rabinovich}
\orcid{0000-0003-3078-6705}
\affiliation{
  \institution{Upwork}
  \city{Seattle}
  \country{United States}
}
\email{andrewrabinovich@upwork.com}

\author{Wenge Rong}
\orcid{0000-0002-4229-7215}
\affiliation{
  \institution{Beihang University}
  \city{Beijing}
  \country{China}
}
\email{w.rong@buaa.edu.cn}

\author{Fuli Feng$^\dag$}
\orcid{0000-0002-5828-9842}
\affiliation{
  \institution{University of Science and Technology of China}
  \city{Hefei}
  \country{China}
}
\email{fulifeng93@gmail.com}

\def\authors{Yimeng Bai, Chang Liu, Yang Zhang, Dingxian Wang, Frank Yang, Andrew Rabinovich, Wenge Rong, Fuli Feng}

\renewcommand{\shortauthors}{Yimeng Bai et al.}

\input{1_abstract}




\maketitle

\input{2_introduction}

\input{3_preliminary}
\input{4_methodology}
\input{5_experiment}
\input{6_related_work}

\input{7_conclusion}

\begin{acks}
This work is supported by the National Natural Science Foundation of China (62272437).
\end{acks}

\bibliographystyle{ACM-Reference-Format}
\balance
\bibliography{8_reference}

\end{document}

%% file: 1_abstract.tex
\begin{abstract}

Generative recommendation is emerging as a transformative paradigm by directly generating recommended items, rather than relying on matching. Building such a system typically involves two key components: (1) optimizing the tokenizer to derive suitable item identifiers, and (2) training the recommender based on those identifiers. Existing approaches often treat these components separately—either sequentially or in alternation—overlooking their interdependence. This separation can lead to misalignment: the tokenizer is trained without direct guidance from the recommendation objective, potentially yielding suboptimal identifiers that degrade recommendation performance.

To address this, we propose \textbf{BLOGER}, a \textbf{B}i-\textbf{L}evel \textbf{O}ptimization for \textbf{GE}nerative \textbf{R}ecommendation framework, which explicitly models the interdependence between the tokenizer and the recommender in a unified optimization process. The lower level trains the recommender using tokenized sequences, while the upper level optimizes the tokenizer based on both the tokenization loss and recommendation loss. We adopt a meta-learning approach to solve this bi-level optimization efficiently, and introduce gradient surgery to mitigate gradient conflicts in the upper-level updates, thereby ensuring that item identifiers are both informative and recommendation-aligned. Extensive experiments on multiple real-world datasets demonstrate that BLOGER consistently outperforms state-of-the-art generative recommendation methods while maintaining practical efficiency with no significant additional computational overhead, effectively bridging the gap between item tokenization and autoregressive generation. We release our code at \url{https://github.com/Ten-Mao/BLOGER}.

\end{abstract}

%% file: 2_introduction.tex
\section{Introduction}\label{section:intro}
Generative recommendation, as a promising next-generation recommendation paradigm to replace the traditional cascade retrieve-and-rank approach, has garnered attention from both industry and academia~\cite{GeneRec, OneRec, TIGER, HSTU, PinRec, kuaishou_survey}. Unlike the discriminative paradigm, which operates through a query-candidate matching method~\cite{SASRec, HGN}, the generative paradigm can directly generate the identifier of the next item in an autoregressive manner, eliminating the need for one-to-one comparisons with candidate items. This capability arises from the combination of an efficient identifier structure and a powerful generative mechanism, which together improve scalability for large item corpora~\cite{TIGER}, reduce susceptibility to feedback loops~\cite{FeedBackLoop}, and facilitate enhanced user personalization~\cite{zhao2025exploring}.

Generally, the generative recommendation paradigm comprises two core components: \textit{item tokenization} and \textit{autoregressive generation}. The former involves optimizing a \textit{tokenizer} to effectively map each item into a sequence of tokens that serve as its identifier. This process can be achieved through hierarchical clustering~\cite{SEATER, EAGER, ColaRec}, residual quantization~\cite{TIGER, LETTER, ETEGRec, TokenRec, UniGRec}, parallel quantization~\cite{MoC, RPG}, or by directly leveraging textual descriptions~\cite{BIGRec, IDGenRec, LLM2BERT4Rec} or collaborative signals~\cite{LLM-RecSys-ID, P5}. The latter entails optimizing a transformer-based \textit{recommender} (\eg T5~\cite{T5}) to autoregressively generate the next item’s identifier based on the tokenized interaction history.

\begin{figure}
    \centering
    \includegraphics[width=0.475\textwidth]{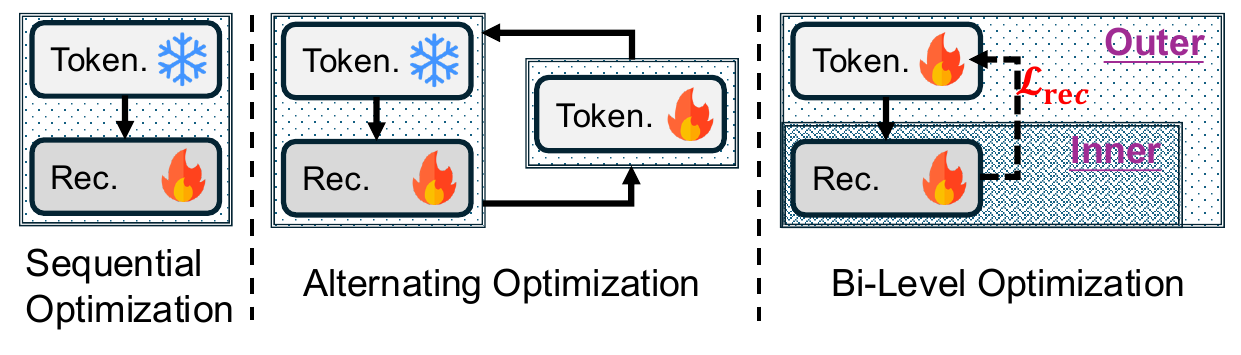}
    \caption{Comparison of optimization strategies in generative recommendation. Snowflake indicates frozen parameters; flame indicates modules being optimized. “Token.” and “Rec.” refer to the tokenizer and recommender, respectively. }
    \label{fig:opt}
    \Description{...}
\end{figure}

Existing works typically treat item tokenization and autoregressive generation as independent components, employing one of two optimization strategies: (1) \textit{sequential optimization}, where the tokenizer is first trained independently to produce static item identifiers, which are then used by the recommender for generation tasks~\cite{TIGER, LETTER, SEATER, EAGER, TokenRec, ColaRec}; or (2) \textit{alternating optimization}, where the tokenizer and recommender are updated in an iterative manner~\cite{IDGenRec, ETEGRec}. However, these decoupled formulations fail to capture the intricate interdependence between tokenization and generation at the optimization level. This often results in misalignment: the tokenizer is trained without direct guidance from the recommendation objective, potentially yielding suboptimal identifiers that impair the recommender's ability to make  predictions. 
\bym{For example, under decoupled formulations, the tokenizer may assign unrelated identifiers to frequently co-purchased items like ``baby formula'' and ``diapers'' due to their modality differences, failing to capture their strong user-behavior correlation and thus limiting the performance of identifier-dependent recommenders.}

To address this misalignment issue, we propose explicitly modeling their mutual dependence within a unified optimization framework. Our key insight is that the tokenizer's efficacy is inherently reflected in the recommender's performance, since superior token representations directly contribute to more precise target item prediction. Conversely, the recommender requires high-quality item identifiers from the tokenizer to effectively model user preferences. Therefore, as illustrated in Figure~\ref{fig:opt}, we reformulate generative recommendation as a \textit{bi-level optimization}~\cite{BLO} problem, where the inner level trains the recommender on tokenized inputs, while the outer level optimizes the tokenizer using both the tokenization loss and the recommendation loss, with the latter providing direct supervision from the downstream recommendation task. This formulation enables explicit alignment between the tokenization and the recommendation objective, fostering mutual reinforcement between the two models.

To this end, we propose the first \textbf{B}i-\textbf{L}evel \textbf{O}ptimization for \textbf{GE}nerative \textbf{R}ecommendation (\textbf{BLOGER}) framework. To efficiently solve the nested optimization structure, we adopt a meta-learning approach~\cite{MAML}, which bridges the inner and outer levels via meta-gradients~\cite{LabelCraft}, thereby enabling more effective bidirectional learning between the tokenizer and the recommender~\cite{EBLO}.
Furthermore, to address potential conflicts between the tokenization and recommendation objectives during outer-level optimization, we incorporate gradient surgery~\cite{PCGrad}. This allows for the precise extraction of the component of the recommendation loss that is beneficial for optimizing the tokenizer, ensuring that the resulting item identifiers align with both tokenization quality and recommendation performance.
We empirically evaluate the framework through extensive experiments on multiple real-world datasets. The results underscore that BLOGER consistently delivers superior performance in generative recommendation tasks.

The main contributions of this work are summarized as follows:
\begin{itemize}[leftmargin=*]
\item We propose reformulating generative recommendation as a bi-level optimization problem, explicitly modeling the mutual dependence between the tokenizer and the recommender.
\item We introduce BLOGER, a framework that \textit{effectively} and \textit{efficiently} solves the bi-level optimization problem through meta-learning and gradient surgery techniques, \bym{without introducing significant additional computational overhead}.
\item We conduct extensive experiments on multiple real-world datasets, demonstrating the effectiveness of our method.
\end{itemize}

%% file: 3_preliminary.tex
\section{Preliminary}\label{section:pre}
In this section, we formally define the task of generative recommendation. In line with prior studies~\cite{TIGER, LETTER, ETEGRec}, we focus on a sequential recommendation setting. Given a universal item set $\mathcal{I}$ and a user's historical interaction sequence $S = [i_1, i_2, ..., i_T]$, the goal is to predict the next item $i_{T+1} \in \mathcal{I}$ that the user is likely to interact with. Departing from the traditional query-candidate matching approach~\cite{SASRec}, generative recommendation approaches reframe the task as a direct item generation problem. This paradigm typically comprises the following two core components:

\vspace{+3pt}
\textbf{Item Tokenization.}
This component assigns an identifier to each item, represented as a sequence of tokens, to facilitate both item encoding and generation. Specifically, the input interaction sequence $S$ can be mapped into a token sequence, denoted as $X=[c_1^1, c_2^1, \ldots,c_{L-1}^{T},c_{L}^T]$, where $c_l^j$ denotes the $l$-th token of item $i_j$ $(j\in\{1,\ldots,T\})$, $T$ is the length of the original interaction sequence $S$, and $L$ is the identifier length. Although $L$ may vary across items, we adopt a fixed-length tokenization scheme to mitigate potential bias in item prediction caused by varying sequence lengths~\cite{D3}. The ground-truth next item $i_{T+1}$ is similarly represented as a token sequence $Y = [c_1^{T+1}, \ldots, c_L^{T+1}]$.

\vspace{+3pt}
\textbf{Autoregressive Generation.}
Given the tokenized interaction history, the next-item prediction task is reformulated as a sequence generation problem. Specifically, the objective is to model the conditional probability of the target token sequence as:
\begin{equation}
P(Y|X)=\prod_{l=1}^{L}P(c_{l}^{T+1}|X,c_1^{T+1},\ldots,c_{l-1}^{T+1}),
\end{equation}
where each token of the target item is generated in an autoregressive manner, conditioned on both the input sequence and the previously generated tokens.

%% file: 4_methodology.tex
\section{Methodology}
\begin{figure}[t]
    \centering
    \includegraphics[width=0.475\textwidth]{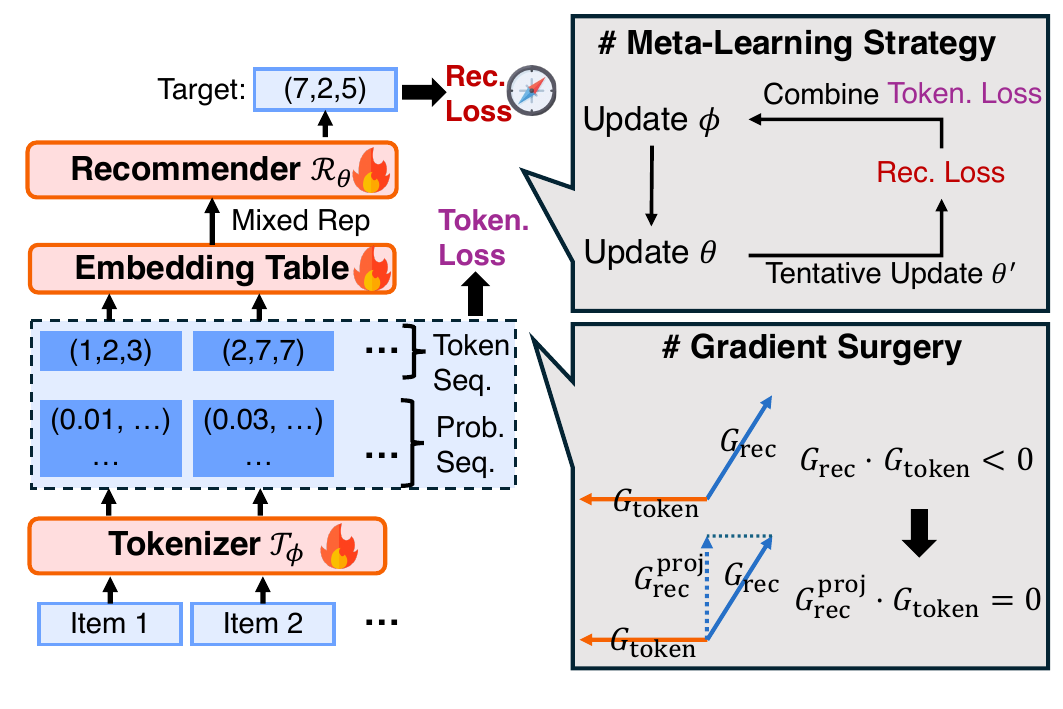}
    \caption{An overview of the proposed BLOGER framework, which comprises an encoder-decoder model architecture and a tailored learning scheme. Specifically, we construct a mixed representation based on the assignment probabilities over the codebook to ensure the differentiability of the recommendation loss to the tokenizer. Besides, we adopt a meta-learning strategy to enable efficient optimization, and apply gradient surgery to extract beneficial components from the gradients of the recommendation loss.}
    \label{fig:framework}
    \Description{...}
\end{figure}

In this section, we first present the formulation of the bi-level optimization problem, followed by the detailed introduction of the proposed BLOGER framework. Finally, we provide a discussion and comparison with existing methods.

\subsection{Problem Reformulation}
Our reformulation is motivated by the inherent dependency between the tokenizer and the recommender in the generative recommendation paradigm. Specifically, the tokenizer's performance directly influences the recommender’s effectiveness, as superior token representations lead to more accurate target item predictions. Conversely, the recommender depends on the tokenizer to provide high-quality item identifiers, which are crucial for accurately modeling user preferences. Given this interdependence, \textit{bi-level optimization}~\cite{EBLO} offers a principled formulation for effectively modeling their interaction.

We denote the tokenizer and the recommender as $\mathcal{T}_\phi$ and $\mathcal{R}_\theta$, respectively, where $\phi$ and $\theta$ are the corresponding model parameters. The bi-level optimization problem is formulated as follows:

\begin{subequations}\label{eq:blo}
\begin{align}
\label{eq:outer}
\min_{\phi} \quad & \mathcal{L}_{\text{rec}}(\mathcal{T}_\phi, \mathcal{R}_{\theta^*}) + \lambda \mathcal{L}_{\text{token}}(\mathcal{T}_\phi) \\ 
\label{eq:inner}
\text{s.t.} \quad & \theta^* = \arg\min_{\theta} \mathcal{L}_{\text{rec}}(\mathcal{T}_\phi, \mathcal{R}_\theta). 
\end{align}
\end{subequations}
Here, the recommendation loss $\mathcal{L}_{\text{rec}}(\mathcal{T}_\phi, \mathcal{R}_\theta)$ measures the performance of the recommender $\mathcal{R}_\theta$ on the training dataset, based on the tokenized outputs generated by the tokenizer $\mathcal{T}_\phi$. The parameter $\theta^*$ represents the optimal recommender parameters obtained by minimizing this recommendation loss and is therefore dependent on $\phi$. Consequently, $\mathcal{L}_{\text{rec}}(\mathcal{T}_\phi, \mathcal{R}_{\theta^*})$ denotes the recommendation loss of the trained recommender $\mathcal{R}_{\theta^*}$, conditioned on the tokenizer $\mathcal{T}_\phi$. The tokenization loss $\mathcal{L}_{\text{token}}(\mathcal{T}_\phi)$ is independent of the recommender and serves to evaluate the intrinsic quality of the tokenizer. The hyperparameter $\lambda$ controls the relative importance of this loss in the overall objective. Detailed definitions of these loss functions will be provided in the following sections.

Clearly, Equation~\eqref{eq:inner} formulates the inner-level objective, which trains the recommender $\mathcal{R}_\theta$ to minimize the recommendation loss under a fixed tokenizer $\mathcal{T}_\phi$. In contrast, Equation~\eqref{eq:outer} defines the outer-level objective, which seeks an optimal tokenizer $\mathcal{T}_\phi$ such that the recommender trained on its tokenized outputs achieves minimal recommendation loss, while the tokenizer simultaneously minimizes its own tokenization loss. This bi-level optimization framework offers a principled formulation of the generative recommendation task, effectively bridging item tokenization and autoregressive generation through the learned recommender $\mathcal{R}_{\theta^*}$.

\subsection{BLOGER Framework}
We provide an overview of the BLOGER framework in Figure~\ref{fig:framework}. In the following, we detail its components, including the architecture of the tokenizer and recommender, and the tailored learning scheme for solving the bi-level optimization problem.
\subsubsection{Tokenizer}
Building on recent advances~\cite{TIGER, LETTER, ETEGRec}, we adopt residual quantization~\cite{RQ} to represent each item as a sequence of $L$ discrete tokens. This hierarchical encoding offers two main benefits. First, it inherently induces a tree-structured item space, which is well-suited for generative modeling. Second, items that share the same prefix tokens can naturally capture collaborative semantics, enabling the sharing of information among related items.

Specifically, we implement the item tokenizer $\mathcal{T}_\phi$ using a RQ-VAE~\cite{RQ} model. Given an item $i$, the tokenizer takes its semantic embedding $\bm{z}$—typically a pretrained text embedding or a collaborative embedding—as input and produces a sequence of quantized tokens across $L$ hierarchical levels, formulated as:
\begin{equation}
[c_1,\ldots, c_L] = \mathcal{T}_\phi(\bm{z}),
\end{equation}
where $c_l$ represents the token assigned to item $i$ at the $l$-th level.

\vspace{+3pt}
\textbf{Quantization.} 
Next, we describe the quantization process in detail. We begin by encoding the input embedding $\bm{z}$ into a latent representation using a MLP-based encoder:
\begin{equation}
\bm{r} = \text{Encoder}_\mathcal{T}(\bm{z}).
\end{equation}
The latent representation $\bm{r}$ is then quantized into $L$ discrete tokens by querying $L$ distinct codebooks. At each level $l$, we have a codebook $\mathcal{C}_{l}=\{\bm{e}_l^k\}_{k=1}^{K}$, where $K$ is the codebook size. The residual quantization process is performed as follows:
\begin{align}
\label{eq:prob}
P(k | \bm{v}_l) &= \frac{\exp\left(-\lVert \bm{v}_l - \bm{e}_l^k \rVert^2\right)}{\sum_{j=1}^{K} \exp\left(-\lVert \bm{v}_l - \bm{e}_l^j \rVert^2\right)}, \\
c_l &= \underset{k}{\arg\max} \; P(k | \bm{v}_l), \\
\bm{v}_l &= \bm{v}_{l-1} - \bm{e}_{l-1}^{c_{l-1}}.
\end{align}
Here, $\bm{v}_l$ denotes the residual vector at the $l$-th level ($\bm{v}_1=\bm{r}$), and $P(k | \bm{v}_l)$ represents the probability of assigning the residual to token $k$, computed based on the Euclidean distance between $\bm{v}_l$ and each entry in the corresponding codebook. 

\vspace{+3pt}
\textbf{Reconstruction.} 
After quantizing the initial semantic embedding into a hierarchy of tokens from coarse to fine granularity, we obtain the quantized representation $\tilde{\bm{r}} = \sum_{l=1}^{L} \bm{e}_{l}^{c_l}$, which is then fed into an MLP-based decoder to reconstruct the semantic embedding:
\begin{equation}
\tilde{\bm{z}} = \text{Decoder}_{\mathcal{T}}(\tilde{\bm{r}}).
\end{equation}

\vspace{+3pt}
\textbf{Tokenization Loss.} 
Overall, the tokenization loss in Equation~\eqref{eq:outer} is defined as:
\begin{equation}\label{eq:loss_token}
\mathcal{L}_\text{token}(\mathcal{T}_\phi) = \sum_i (\lVert \tilde{\bm{z}} - \bm{z} \rVert^2 + \sum_{l=1}^{L} \lVert \text{sg}[\bm{v}_l] - \bm{e}_l^{c_l} \rVert^2 + \beta \lVert \bm{v}_l - \text{sg}[\bm{e}_l^{c_l}] \rVert^2),
\end{equation}
where the loss is computed for each item $i$ in the sequence and then summed. The first term is a reconstruction loss that encourages the reconstructed embedding $\tilde{\bm{z}}$ to closely approximate the original input $\bm{z}$. The remaining two terms are quantization losses that minimize the discrepancy between the residual vectors and the selected codebook entries. Here, $\text{sg}[\cdot]$ denotes the stop-gradient operation~\cite{RQ}, and $\beta$ is a weighting coefficient, typically set to 0.25, that balances the updates between the encoder and the codebooks.

\subsubsection{Recommender}
For the generative recommender $\mathcal{R}_\theta$, we adopt a Transformer-based encoder-decoder architecture T5~\cite{T5}, which has demonstrated strong performance in prior generative recommendation research~\cite{TIGER,LETTER}. 

\vspace{+3pt}
\textbf{Mixed Representation.} 
Given the tokenizer $\mathcal{T}_\phi$, the user’s item-level interaction sequence $S = [i_1, \ldots, i_T]$ and the target item $i_{T+1}$ are respectively tokenized into the token-level input sequence $X = [c_1^1, c_2^1, \ldots, c_{L-1}^{T}, c_{L}^T]$ and the target sequence $Y = [c_1^{T+1}, \ldots, c_L^{T+1}]$. However, since the tokenizer outputs discrete token sequences, the process is inherently non-differentiable, making it challenging to directly optimize the tokenizer using the outer-level recommendation loss in Equation~\eqref{eq:outer}.

To address this issue, we leverage differentiable soft representations to facilitate gradient-based optimization. Let $\bm{E}^V$ denote the vocabulary embeddings of the recommender $\mathcal{R}_\theta$, where each row corresponds to the embedding of a token in the vocabulary. For a given token $c_l^t$ (where $l = 1, \ldots, L$ and $t = 1, \ldots, T+1$), we first obtain its \textit{hard embedding} $\bm{h}_l^t$ by directly looking up the corresponding vector in the vocabulary embeddings $\bm{E}^V$. In parallel, we pad the probability distribution $P(k | \bm{v}_l^t)$—derived from the corresponding codebook via Equation~\eqref{eq:prob}—with zeros to match the size of the full vocabulary. Using this padded distribution, we compute the \textit{soft embedding} $\bm{s}_l^t$ as a weighted average over the vocabulary embeddings $\bm{E}^V$, with weights corresponding to the padded probabilities. Finally, the mixed representation of the token is computed as:
\begin{equation}
\bm{o}_l^t=\bm{s}_l^t+\text{sg}[\bm{h}_l^t-\bm{s}_l^t],
\end{equation}
where $\text{sg}[\cdot]$ denotes the stop-gradient operation. This mixed representation not only preserves the numerical consistency of the hard embedding but also enables end-to-end differentiability by allowing gradients to flow exclusively through the soft embedding.

\vspace{+3pt}
\textbf{Seq2Seq Formulation.}
During training, we first compute the mixed representations for all tokens in $X$ to form the input sequence embedding $\bm{E}^X = [\bm{o}_1^1, \bm{o}_2^1, \ldots, \bm{o}_{L-1}^T, \bm{o}_L^T]$. It is then fed into the encoder to obtain the corresponding hidden representation:
\begin{equation}
\bm{H}^{E}=\text{Encoder}_{\mathcal{R}}(\bm{E}^X).
\end{equation}
For decoding, we prepend a special beginning-of-sequence token [BOS] to the target sequence $Y$, and compute its mixed representation  $\bm{E}^Y=[\bm{o}^{\text{BOS}}, \bm{o}_1^{T+1},\ldots,\bm{o}_{L}^{T+1}]$. The decoder then takes the encoder output $\bm{H}^E$ and the target embedding sequence $\bm{E}^Y$ as input to model user preference representation:
\begin{equation}
\bm{H}^{D}=\text{Decoder}_{\mathcal{R}}(\bm{H}^{E},\bm{E}^Y).
\end{equation}

\vspace{+3pt}
\textbf{Recommendation Loss.} 
The decoder output $\bm{H}^{D}$ is projected onto the vocabulary space through an inner product with the vocabulary embedding matrix $\bm{E}^V$ to predict the target item tokens. The recommendation loss in Equation~\eqref{eq:blo} involves optimizing the negative log-likelihood of the target tokens following the sequence-to-sequence learning paradigm:
\begin{equation}\label{eq:loss_rec}
\mathcal{L}_\text{rec}(\mathcal{T}_\phi,\mathcal{R}_\theta)=-\sum_{l=1}^{L}\log P(Y_l|X,Y_{<l}),
\end{equation}
where $Y_l$ denotes the $l$-th token in the target sequence, and $Y_{<l}$ represents all preceding tokens. This formulation enables the model to generate the target item tokens in an autoregressive manner.

\subsubsection{Tailored Learning Scheme}
After introducing the model architecture, we now describe the approach for solving the bi-level optimization problem defined in Equation~\eqref{eq:blo}.

\vspace{+3pt}
\textbf{Meta-Learning Strategy.} 
The two levels of optimization are mutually dependent, resulting in a nested optimization loop that is inherently difficult to solve directly. To address this challenge, we propose a training strategy based on the meta-learning algorithm MAML~\cite{MAML}, which iteratively updates $\mathcal{T}_\phi$ and $\mathcal{R}_\theta$ in an alternating fashion. Specifically, the optimization proceeds as follows:

\begin{itemize}[leftmargin=*]
\item 
\textbf{Update of $\theta$.}
For the inner-level objective, we directly apply tokenization and optimize the recommendation loss to update $\theta$. This update is formulated as:
\begin{equation}\label{eq:update_theta}
\theta \leftarrow \theta -\eta_{\mathcal{R}}\nabla_\theta\mathcal{L}_\text{rec}(\mathcal{T}_\phi,\mathcal{R}_\theta),
\end{equation}
where $\eta_{\mathcal{R}}$ represents the learning rate for the recommender.

\item 
\vspace{+3pt}
\textbf{Update of $\phi$.}
For the outer-level objective, updating the tokenizer $\mathcal{T}_\phi$ requires obtaining the optimal recommender parameters $\theta^*$, which involves fully training the recommender model — a process that is computationally prohibitive. Therefore, we follow prior meta-learning approaches and perform a tentative update of $\mathcal{R}_\theta$ instead. The performance of this temporarily updated recommender is then used to guide the optimization of the tokenizer. Concretely, it consists of two steps:
\begin{itemize}[leftmargin=*]
\vspace{+3pt}
\item[-] \textit{Meta training}. First, we keep the tokenizer unchanged and fictively update the recommender. In particular, we fix the parameter $\phi$ and compute the recommendation loss in Equation~\eqref{eq:loss_rec}. Then, a step of gradient descent is performed to obtain a tentative recommender $\mathcal{R}_{\theta^\prime}$. Formally, we have:
\begin{equation}\label{eq:update_phi_1}
\theta^\prime=\theta-\eta_{\mathcal{R}}\nabla_\theta\mathcal{L}_\text{rec}(\mathcal{T}_\phi,\mathcal{R}_\theta).
\end{equation}

\vspace{+3pt}
\item[-] \textit{Meta testing}.
After obtaining the tentative recommender $\mathcal{R}_{\theta^\prime}$, we use it to compute the recommendation loss $\mathcal{L}_\text{rec}(\mathcal{T}_\phi, \mathcal{R}_{\theta^\prime})$. In addition, we compute the tokenization loss $\mathcal{L}_\text{token}(\mathcal{T}_\phi)$. The combination of both losses constitutes the overall outer-level loss. We then update $\phi$ as follows:
\begin{equation}\label{eq:update_phi_2}
\phi \leftarrow \phi - \eta_{\mathcal{T}} 
(\nabla_\phi\mathcal{L}_\text{rec}(\mathcal{T}_\phi,\mathcal{R}_{\theta^\prime})+\lambda\nabla_\phi \mathcal{L}_\text{token}(\mathcal{T}_\phi)),
\end{equation}
where $\eta_{\mathcal{T}}$ represents the learning rate for the tokenizer. The meta-gradient of recommendation loss here can be computed by using the back-propagation along the chain: $$\nabla_\phi\mathcal{L}_\text{rec}(\mathcal{T}_\phi,\mathcal{R}_{\theta^\prime})\rightarrow \theta^\prime \rightarrow \nabla_\theta\mathcal{L}_\text{rec}(\mathcal{T}_\phi,\mathcal{R}_\theta)\rightarrow \phi.$$
\end{itemize}

\end{itemize}
The two updates described above are iteratively applied until convergence. Although this alternating optimization strategy does not guarantee convergence to a global optimum, it has been shown to work well empirically for bi-level optimization problems~\cite{MAML}.

\vspace{+3pt}

\textbf{Gradient Surgery.} 
In practice, we have observed that the cosine similarity between the two gradients in Equation~\eqref{eq:update_phi_2} is often negative\footnote{Notably, during optimization, we find that the proportion of modules exhibiting conflicting gradients can reach as high as 90\%, highlighting the severity of this issue.}. This phenomenon, referred to as gradient conflict~\cite{PCGrad}, can negatively impact the performance of multi-task learning. To address this issue and simultaneously optimize the tokenization loss and recommendation loss at the outer level, we apply gradient surgery~\cite{PCGrad,GradCraft} to resolve conflicts at the gradient level.

The key idea is that our goal is to retain the beneficial contribution of the recommendation loss while eliminating only the part that conflicts with the tokenization objective. Specifically, when the cosine similarity between the two gradients in Equation~\eqref{eq:update_phi_2} is negative, we project the gradient of the recommendation loss onto the normal plane of the tokenization loss gradient to remove the conflicting component.  Formally, this adjustment is expressed as:
\begin{equation}\label{eq:pcgrad}
\begin{aligned}
G_\text{rec}^{\text{proj}} &= G_\text{rec} - \frac{G_\text{rec} \cdot G_\text{token}}{\Vert G_\text{token} \Vert^2} G_\text{token}, \\
G_\text{rec} &= \nabla_\phi \mathcal{L}_\text{rec}(\mathcal{T}_\phi, \mathcal{R}_{\theta^\prime}), \\
G_\text{token} &= \nabla_\phi \mathcal{L}_\text{token}(\mathcal{T}_\phi).
\end{aligned}
\end{equation}
Here, $G_\text{rec}$ and $G_\text{token}$ denote the two gradients in Equation~\eqref{eq:update_phi_2}. After the adjustment, the inner product satisfies $G_\text{rec}^{\text{proj}} \cdot G_\text{token} = 0$, implying that the cosine similarity between them becomes zero. By substituting the original gradient $G_\text{rec}$ with its orthogonal projection $G_\text{rec}^{\text{proj}}$, this approach effectively eliminates the conflicting component, thereby mitigating gradient interference and enhancing the stability of multi-task optimization. Importantly, the above operation is performed separately for each named parameter group, instead of applying the projection to the model parameters as a whole, enabling finer-grained conflict mitigation across different parts of the network.

\subsubsection{Complexity Analysis}\label{sec:comp}

Algorithm~\ref{alg:training} outlines the detailed training procedure. In each iteration, the algorithm first updates the recommender parameters $\theta$ (Lines 2–8), followed by an update of the tokenizer parameters $\phi$ (Line 9), which occurs every $M$ training steps. This delay in updating $\phi$ allows the recommender to optimize more thoroughly before providing stable gradients for the tokenizer.
In practice, both updates are performed on mini-batches of data to ensure computational efficiency.

We now analyze the computational complexity of the proposed framework. Let $d$ denote the model dimension, $K$ the size of each codebook, $L$ the number of codebooks, and $T$ the sequence length. For item tokenization, considering a single item as an example, the time complexity of the encoder and decoder layers is $\mathcal{O}(d^2)$. 
Obtaining the mixed representation—comprising the codebook lookup and soft representation operations—incurs an additional computational complexity of $\mathcal{O}(LKd)$. The computation of the tokenization loss incurs a cost of $\mathcal{O}(d + Ld)$. Therefore, the overall time complexity for tokenizing one item is $\mathcal{O}(d^2 + LKd)$.
For generative recommendation, the primary computational cost arises from the self-attention and feed-forward layers, which have a complexity of $\mathcal{O}(T^2d + Td^2)$. Additionally, computing the recommendation loss incurs a complexity of $\mathcal{O}(TLKd)$.
Although the meta-learning strategy and gradient surgery slightly increase the number of forward and backward passes, they do not alter the overall order of computational complexity. 

In summary, the total training cost of BLOGER is $\mathcal{O}(TLKd + T^2d + Td^2)$, matching the order of magnitude of TIGER~\cite{TIGER}, while its inference complexity remains identical. \textbf{Theoretically, this indicates that the operation of BLOGER maintains practical efficiency without introducing any significant additional computational overhead.}

\subsection{Discussion}

To emphasize the methodological innovations of our proposed framework, we compare it with existing approaches in terms of optimization formulation, alignment with recommendation objectives, and token sequence generation for the recommender.

As shown in Table~\ref{table:comparison}, existing methods such as TIGER and LETTER rely on a sequential optimization strategy and use static, pre-processed token sequences, which lack awareness of the recommendation task. ETEGRec introduces alternating optimization and gradually refines token sequences by constructing auxiliary losses to align the intermediate layers of the tokenizer and the recommender. However, it requires additional adapter layers and does not directly align with the recommendation objective. 

In contrast to previous approaches, BLOGER introduces the following key innovations: (1) \textbf{Bi-Level Optimization Formulation}: BLOGER adopts a bi-level optimization framework that more accurately captures the interdependence between the tokenizer and the recommender. (2) \textbf{Recommendation-Aligned Tokenization}: The tokenizer is directly optimized with respect to the recommendation loss, eliminating the need for handcrafted auxiliary objectives and ensuring that the tokenization process is tightly aligned with the end task. (3) \textbf{Meta-Gradient-Based Coupling}: 
BLOGER leverages meta-gradients~\cite{LabelCraft} to establish a \textit{gradient-level} connection between the recommender and the tokenizer, allowing the tokenizer to directly receive optimization signals from the recommender. Compared to naive joint learning, which updates both modules simultaneously without explicitly modeling their interdependence, this approach achieves more effective coordination. Furthermore, unlike \textit{iteration-independent}~\cite{EBLO} alternating updates used in ETEGRec, gradient-level coupling capture the interactions between the two modules at a finer granularity.

\begin{algorithm}[t]
\caption{Training of BLOGER}
\label{alg:training}
\begin{algorithmic}[1]
\While{not converged}
\State Update $\theta$ according to Equation~\eqref{eq:update_theta};
\If{\text{Step} \% $M == 0$}

\State Compute $\theta^\prime$ with Equation~\eqref{eq:update_phi_1};
\State Compute $G_\text{rec}$ and $G_\text{token}$;
\If{$G_\text{rec}\cdot G_\text{token}<0$}
\State Compute $G_\text{rec}^{\text{proj}}$ with Equation~\eqref{eq:pcgrad};
\State $G_\text{rec} \leftarrow G_\text{rec}^{\text{proj}}$
\EndIf
\State Update $\phi$ according to Equation~\eqref{eq:update_phi_2};
\EndIf
\EndWhile
\end{algorithmic}
\end{algorithm}

\begin{table}[t]
\caption{Comparison of BLOGER with several related generative recommendation methods.}
\label{table:comparison}
\resizebox{\columnwidth}{!}{
\begin{tabular}{cccc}
\hline
Method  & Optimization & Alignment      & Token Sequence    \\ \hline
TIGER~\cite{TIGER}   & Sequential   & \ding{55}   & Pre-processed     \\
LETTER~\cite{LETTER}  & Sequential   & \ding{55}   & Pre-processed     \\
ETEGRec~\cite{ETEGRec} & Alternating  & Auxiliary Loss & Gradually Refined \\
BLOGER  & Bi-Level     & Rec Loss       & Gradually Refined \\ \hline
\end{tabular}
}
\end{table}

%% file: 5_experiment.tex
\section{Experiment}

In this section, we conduct a series of experiments to answer the
following research questions:

\noindent \textbf{RQ1}: How does BLOGER perform compared to exist generative recommendation methods?

\noindent \textbf{RQ2}: What is the contribution of each individual component to the overall effectiveness of BLOGER?

\noindent \textbf{RQ3}: How do specific hyper-parameters influence BLOGER?

\noindent \textbf{RQ4}: How is the time efficiency of BLOGER?

\noindent \textbf{RQ5}: What are the underlying factors driving the performance gains of BLOGER?

\subsection{Experimental Setting}

\subsubsection{Datasets}
We conduct experiments on three subsets of the Amazon Review Data~\cite{Amazon14,Amazon18}, namely \textit{Beauty}, \textit{Instruments}, and \textit{Arts}\footnote{\url{https://nijianmo.github.io/amazon/index.html}} for evaluation. The dataset statistics are presented in Table~\ref{table:dataset}. Following established practices, we apply the 5-core filtering strategy~\cite{SASRec,TIGER}, removing users and items with fewer than five interactions. Each user's interaction history is truncated or padded to a fixed length of 20. We adopt the widely used \textit{leave-one-out} strategy to split the data into training, validation, and test sets. Specifically, for each user, the latest interaction is used as testing data, the second most recent interaction is validation data, and all other interaction records are used for training.

\begin{table}[t]
\caption{Statistical details of the evaluation datasets, where “AvgLen” represents the average length of item sequences. }
\label{table:dataset}
\resizebox{\columnwidth}{!}{
\begin{tabular}{cccccc}
\hline
Dataset     & \multicolumn{1}{l}{\#User} & \multicolumn{1}{l}{\#Item} & \multicolumn{1}{l}{\#Interaction} & \multicolumn{1}{l}{Sparsity} & \multicolumn{1}{l}{AvgLen} \\ \hline
Beauty      & 22363                      & 12101                      & 198502                            & 99.93\%                      & 8.87                       \\
Instruments & 24772                      & 9922                       & 206153                            & 99.92\%                      & 8.32      \\ 
Arts & 45141                      & 20956                       & 390832                            & 99.96\%                      & 8.66                       \\ \hline
\end{tabular}
}
\end{table}

\begin{table*}[t]
\caption{The overall performance comparisons between the baselines and BLOGER. Recall@$K$ and NDCG@$K$ are abbreviated as R@$K$ and N@$K$, respectively. The best and second-best results are highlighted in bold and underlined font, respectively.}
\label{table:main}
\resizebox{1.0\linewidth}{!}{
\begin{tabular}{cccccccccccccc}
\hline
Dataset                      & Metric & MF                         & LightGCN & Caser  & GRU4Rec & HGN    & SASRec & P5                 & TIGER  & OneRec & LETTER             & ETEGRec            & BLOGER          \\ \hline
\multirow{4}{*}{Beauty}      & R@5    & 0.0226                     & 0.0208   & 0.0110 & 0.0229  & 0.0381 & 0.0372 & 0.0400             & 0.0384 & 0.0408 & \underline{0.0424} & 0.0413             & \textbf{0.0444} \\
                             & N@5    & 0.0142                     & 0.0127   & 0.0072 & 0.0155  & 0.0241 & 0.0237 & \underline{0.0274} & 0.0256 & 0.0258 & 0.0264             & 0.0271             & \textbf{0.0293} \\
                             & R@10   & 0.0389                     & 0.0351   & 0.0186 & 0.0327  & 0.0558 & 0.0602 & 0.0590             & 0.0609 & 0.0621 & 0.0612             & \underline{0.0632} & \textbf{0.0654} \\
                             & N@10   & 0.0195                     & 0.0174   & 0.0096 & 0.0186  & 0.0297 & 0.0311 & 0.0335             & 0.0329 & 0.0332 & 0.0334             & \underline{0.0342} & \textbf{0.0361} \\ \hline
\multirow{4}{*}{Instruments} & R@5    & 0.0456                     & 0.0722   & 0.0425 & 0.0559  & 0.0797 & 0.0709 & 0.0809             & 0.0865 & \underline{0.0879} & 0.0872             & 0.0878 & \textbf{0.0882} \\
                             & N@5    & 0.0376                     & 0.0608   & 0.0348 & 0.0459  & 0.0676 & 0.0565 & 0.0695             & 0.0736 & 0.0743 & \underline{0.0747} & 0.0745             & \textbf{0.0753} \\
                             & R@10   & 0.0511                     & 0.0887   & 0.0528 & 0.0702  & 0.0967 & 0.0922 & 0.0987             & 0.1062 & 0.1073 & \underline{0.1082} & 0.1079             & \textbf{0.1100} \\
                             & N@10   & 0.0394                     & 0.0661   & 0.0381 & 0.0505  & 0.0731 & 0.0633 & 0.0751             & 0.0799 & 0.0801 & \underline{0.0814} & 0.0810             & \textbf{0.0822} \\ \hline
\multirow{4}{*}{Arts}        & R@5    & 0.0473                     & 0.0464   & 0.0571 & 0.0669  & 0.0667 & 0.0758 & 0.0724             & 0.0807 & 0.0828 & 0.0841             & \underline{0.0860} & \textbf{0.0879} \\
                             & N@5    & 0.0271                     & 0.0244   & 0.0407 & 0.0518  & 0.0516 & 0.0632 & 0.0607             & 0.0640 & 0.0653 & 0.0675             & \underline{0.0687} & \textbf{0.0703} \\
                             & R@10   & \multicolumn{1}{l}{0.0753} & 0.0755   & 0.0781 & 0.0834  & 0.0910 & 0.0945 & 0.0902             & 0.1017 & 0.1067 & 0.1081             & \underline{0.1084} & \textbf{0.1108} \\
                             & N@10   & \multicolumn{1}{l}{0.0361} & 0.0338   & 0.0474 & 0.0523  & 0.0595 & 0.0693 & 0.0664             & 0.0707 & 0.0730 & 0.0752             & \underline{0.0759} & \textbf{0.0777} \\ \hline
\end{tabular}
}
\vspace{+3pt}
\end{table*}

\subsubsection{Baselines}
The baseline models used for comparison fall into the following two categories:

\vspace{+3pt}
\noindent (1) \textit{Traditional recommender models}:
\begin{itemize}[leftmargin=*]
\item \textbf{MF}~\cite{MF} decomposes the user-item interactions into the user embeddings and the item embeddings in the latent space.

\item \textbf{LightGCN}~\cite{LightGCN} captures high-order user-item interactions through a lightweight graph convolutional network.

\item \textbf{Caser}~\cite{Caser} utilizes horizontal and vertical convolutional filters to model user behavior sequences.

\item \textbf{GRU4Rec}~\cite{GRU4Rec} is an RNN-based sequential recommender, which utilizes GRU to encode historical interactions.

\item \textbf{HGN}~\cite{HGN} employs hierarchical gating networks to capture both long-term and short-term user interests from item sequences.

\item \textbf{SASRec}~\cite{SASRec} employs self-attention mechanisms to capture long-term dependencies in user interaction history.

\end{itemize}

\vspace{+3pt}
\noindent (2) \textit{Generative recommender models}:
\begin{itemize}[leftmargin=*]
\item \textbf{P5}~\cite{P5} integrates collaborative knowledge into LLM-based generative recommender by generating item identifiers through spectral clustering on item co-occurrence graphs.

\item \textbf{TIGER}~\cite{TIGER} employs an RQ-VAE to encode item representations into discrete semantic IDs, and further adopts a generative retrieval paradigm for sequential recommendation.

\item \textbf{OneRec}~\cite{deng2025onerec} leverages RQ-Kmeans to learn hierarchical indexing.

\item \textbf{LETTER}~\cite{LETTER} enhances TIGER by introducing a learnable tokenizer that incorporates hierarchical semantics, collaborative signals, and promotes diversity in code assignment.
 
\item \textbf{ETEGRec}~\cite{ETEGRec} introduces two auxiliary losses to align the intermediate representations of the tokenizer and the recommender, and adopts an alternating optimization strategy to facilitate end-to-end training of both components.

\end{itemize}

\subsubsection{Evaluation Metrics}
To evaluate the performance of various methods in sequential recommendation, we use two widely adopted metrics: Recall@$K$ and NDCG@$K$ ($K=5, 10$). We perform the full ranking evaluation over the entire item set to prevent any bias introduced by sampling. During autoregressive generation, we apply constrained decoding~\cite{LLM-RecSys-ID} using the Trie—a prefix tree that ensures only valid item identifiers are generated. Following prior work~\cite{LETTER}, the beam size is consistently set to 20.

\subsubsection{Implementation Details}
For traditional methods, we follow the implementation provided by~\cite{S3-Rec}. For generative methods, we adopt a unified recommender configuration to ensure a fair comparison. Specifically, we use 4 layers for both the transformer-based encoder and decoder, with model dimensions of 128 and a dropout rate of 0.1. Each layer contains 6 attention heads with a dimension of 64, and the MLP hidden dimension is set to 1024, with ReLU activation applied throughout. For item tokenization, we use 4-level codebooks for RQ-VAE with each codebook initialized via k-means and comprising 256 code embeddings of dimension 32. ETEGRec employs pretrained SASRec embeddings as semantic embeddings, following the original implementation\footnote{We also experimented with text embeddings; however, the results were suboptimal, possibly due to the alignment loss being reliant on collaborative information.}. Other methods utilize LLaMA2-7B~\cite{llama2} to process the title and description, generating text embeddings.

For the proposed BLOGER, we pretrain the tokenizer (codebooks initialized via k-means) for 20,000 epochs with a batch size of 1,024, using the AdamW~\cite{AdamW} optimizer with a learning rate of 1$e$-3 and a weight decay of 1$e$-4. For the subsequent bi-level optimization, we use a batch size of 256, the AdamW optimizer, and early stopping with a patience of 20 based on validation recommendation loss. The learning rates for the generative recommender
and item tokenizer are set to 5$e$-4 and 1$e$-4, respectively. The hyper-parameter $\lambda$ which controls the loss weight in Equation~\eqref{eq:update_phi_2}, is tuned over the set $\{$1, 5$e$-1, 1$e$-1, 1$e$-2, 1$e$-3, 1$e$-4$\}$. For the hyper-parameter $M$ in Algorithm~\ref{alg:training}, it is tuned to ensure that the update of $\phi$ occurs $\{1, 2, ..., 10\}$ times per epoch. For the specific hyper-parameters of the baseline methods, we search the ranges as defined in their respective papers.

\subsection{Overall Performance (RQ1)}

We begin by assessing the overall performance of the compared methods on these datasets. The summarized results are presented in Table~\ref{table:main}, yielding the following observations:
\begin{itemize}[leftmargin=*]

\item BLOGER demonstrates better recommendation performance compared to the baselines on all datasets, which can be attributed to the effectiveness of its bi-level optimization formulation and tailored learning strategy. Importantly, BLOGER does not introduce any additional auxiliary alignment loss, suggesting that the recommendation loss alone provides direct and effective supervision for guiding the tokenization process—thereby validating our intuition regarding the interdependence between tokenization and generation.

\item Traditional recommendation methods generally underperform compared to generative ones, primarily due to their reliance on discrete ID representations, which limits their ability to capture hierarchical semantic structures across multiple levels of granularity. In contrast, generative approaches that incorporate RQ-VAE or RQ-Kmeans can effectively model such semantics, enabling more nuanced differentiation between semantically similar items by leveraging fine-grained representations.

\item Both ETEGRec and LETTER achieve performance improvements over TIGER, which can be attributed to the enhanced alignment between the tokenizer and the recommender. Specifically, LETTER introduces collaborative embeddings as auxiliary signals during the tokenization phase, effectively addressing the challenge of reconciling items' similar semantics with their dissimilar interactions. In contrast, ETEGRec implicitly enables the two models to complement each other's tasks by incorporating auxiliary alignment losses on intermediate layers throughout the alternating optimization process.
\end{itemize}

\subsection{Ablation Study (RQ2)}
To enhance the performance of generative recommendation in BLOGER, we propose the integration of a bi-level optimization formulation along with a corresponding learning scheme. To validate the rationale behind these design decisions, we conduct a thorough evaluation by systematically disabling each critical design element, resulting in a set of variants. Specifically, we introduce the following model variants for comparison:
\begin{itemize}[leftmargin=*]
\item \textbf{w/ LETTER.}
This variant incorporate LETTER into BLOGER, utilizing the pretrained LETTER as the tokenizer and using its collaborative regularization to enhance the learning process.
\item \textbf{w/o GS.}
This variant removes the gradient surgery operation in Equation~\eqref{eq:pcgrad} from the learning strategy, directly using the original gradient without any modifications.
\item \textbf{Joint.}
This variant removes the bi-level optimization formulation, instead combining the recommendation loss and tokenization loss into a single joint optimization for both models.
\item \textbf{Joint w/ GS.}
This variant applies the gradient surgery operation on top of the joint optimization.
\item \textbf{Fixed.}
This variant fixes the tokenizer and optimizes only the recommender, which is functionally equivalent to the design adopted in TIGER.
\end{itemize}
Table~\ref{table:abl} illustrates the comparison results on Beauty, from which
we draw the following observations:
\begin{itemize}[leftmargin=*]
\item Integrating our method with LETTER can further improve recommendation performance compared to using either BLOGER or LETTER alone, which can be attributed to the effectiveness of collaborative regularization in the tokenization stage when text embeddings serve as the quantization basis. This result suggests that incorporating BLOGER with more advanced tokenizers can further enhance performance, thereby confirming the generalizability of our proposed BLOGER framework across diverse backbone methods.

\item Removing the gradient surgery results in a notable performance degradation, suggesting that naively combining the recommendation and tokenization losses without resolving their gradient conflict fails to provide effective supervision for tokenization from the recommendation objective. This underscores the necessity of our method in extracting informative and conflict-free gradients for improved multi-task learning.

\item Joint optimization of the tokenizer and the recommender outperforms the setting with a fixed tokenizer, as it allows the tokenizer to gradually refine the identifier structure during training. Incorporating gradient surgery does not bring significant gains, potentially due to the absence of meta-gradient signals needed to coordinate the optimization between the two components. Nevertheless, the performance still lags behind BLOGER, underscoring the effectiveness of bi-level optimization in capturing the interdependence between tokenization and recommendation.
\end{itemize}

\begin{table}[t]
\caption{Results of the ablation study for BLOGER on Beauty.}
\label{table:abl}\resizebox{1.0\linewidth}{!}{
\begin{tabular}{ccccc}
\hline
Method      & Recall@5 & NDCG@5 & Recall@10 & NDCG@10 \\ \hline
BLOGER      & 0.0444   & 0.0293 & 0.0654    & 0.0361  \\ \hline
w/ LETTER   & \textbf{0.0489}   & \textbf{0.0322} & \textbf{0.0732}    & \textbf{0.0400}  \\
w/o GS      & 0.0329   & 0.0209 & 0.0530    & 0.0274  \\ \hline
Joint       & 0.0418   & 0.0272 & 0.0637    & 0.0342  \\
Joint w/ GS & 0.0412   & 0.0273 & 0.0631    & 0.0343  \\
Fixed       & 0.0384   & 0.0256 & 0.0609    & 0.0329  \\ \hline
\end{tabular}}
\end{table}

\subsection{Hyper-Parameter Analysis (RQ3)}
\begin{figure}[t]
    \centering
    \includegraphics[width=0.475\textwidth]{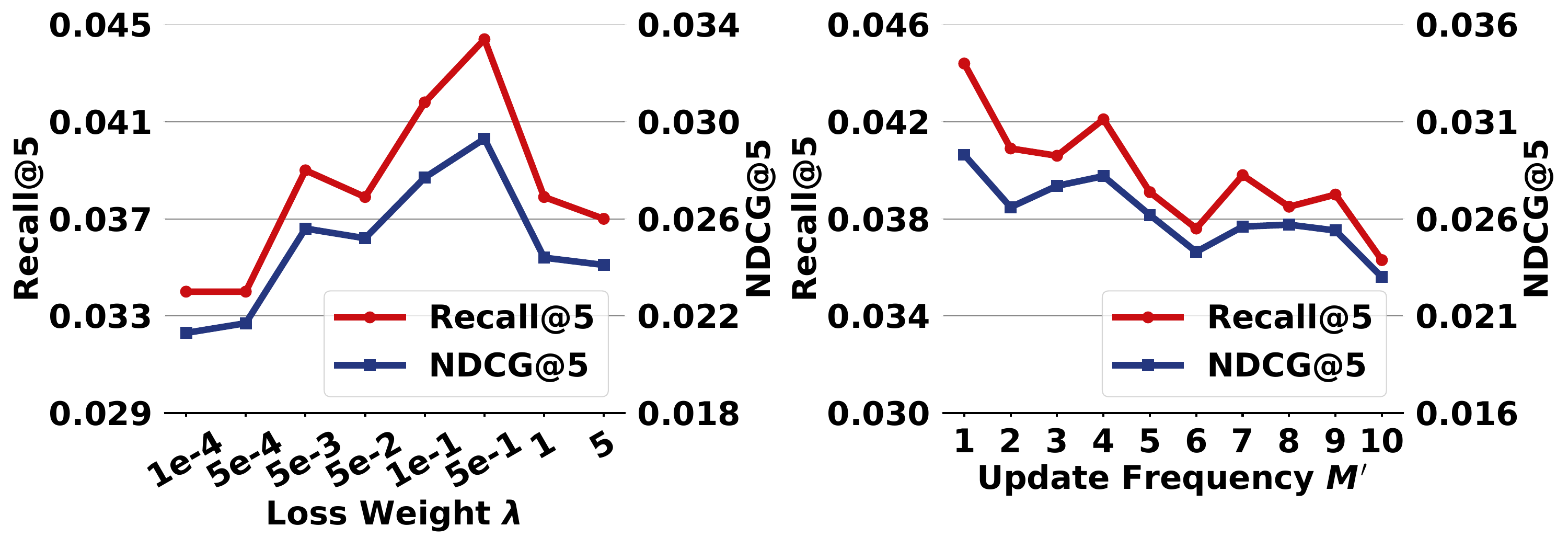}
    \caption{Results of the performance of BLOGER across different values of hyper-parameters. $\lambda$ controls the balance between the recommendation loss and tokenization loss in the outer level, and $M^\prime$ representing the update frequency of the tokenizer parameters $\phi$. }
    \label{fig:hyper}
    \Description{...}
\end{figure}

In our investigation, the two hyper-parameters $\lambda$ and $M$ assume
pivotal roles in influencing the effectiveness. Specifically, $\lambda$ controls the balance between the recommendation loss and tokenization loss in the outer level, $M$ governs the update cycle of the tokenizer. We undertake a systematic examination to scrutinize the impact of varying them on the performance. For $\lambda$, we set it in the range of $\{$1$e$-4, 5$e$-4, 5$e$-3, 5$e$-2, 1$e$-1, 5$e$-1, 1, 5$\}$. For $M$, instead of directly adjusting its absolute value, we introduce a new variable $M'$ to control the number of updates of $\phi$ per epoch, with $M'$ set to $\{1, 2, \dots, 10\}$ per epoch. We summarize the results in Figure~\ref{fig:hyper}, from which we have following observations:
\begin{itemize}[leftmargin=*]
\item When the value of $\lambda$ in Equation~\eqref{eq:update_phi_1} is too small, the constraint imposed by the tokenization loss becomes too weak, potentially disrupting the originally learned identifier structure. Conversely, if $\lambda$ is too large, the recommendation loss is unable to effectively supervise the tokenizer, preventing the desired direct alignment. Empirical evidence suggests that $\lambda = 0.5$ strikes a balance between preserving the identifier structure and enabling effective supervision from the recommendation loss.

\item The update frequency $M^\prime$ of the tokenizer exhibits stable performance when set between 1 and 4, as this delay allows the recommender to optimize more thoroughly before providing stable gradients. In contrast, when the update frequency $M^\prime$ is set too high, the recommender lacks sufficient opportunity to optimize between tokenizer updates, leading to unreliable meta-gradients that undermine the effectiveness of tokenizer training and ultimately degrade overall performance.
\end{itemize}

\subsection{In-Depth Analysis (RQ4 \& RQ5)}

\begin{table}[t]
\caption{Results of the time efficiency comparison between BLOGER and TIGER, where the average training and testing time per epoch are reported.}
\label{table:time}
\resizebox{0.35\textwidth}{!}{
\begin{tabular}{ccc}
\hline
Method & Train (s/epoch) & Test (s/epoch) \\ \hline
TIGER  &   23.7                       &       66.7                            \\
BLOGER &  26.6                       &       66.9                             \\ \hline
\end{tabular}
}
\end{table}
\textbf{Time Efficiency.}
To rigorously assess the time efficiency of our method, we conduct a comparative evaluation of the average training and testing time per epoch for both BLOGER and TIGER under identical GPU configurations (NVIDIA RTX 3090). 
The detailed results are summarized in Table~\ref{table:time}. We observe a slight increase in training time, while the testing time remains nearly identical. Notably, both methods employ the greedy decoding strategy during testing to ensure a fair comparison. These findings suggest that our bi-level optimization learning strategy introduces minimal additional overhead during inference, thereby validating the complexity analysis presented in Section~\ref{sec:comp}.

\vspace{+3pt}
\textbf{Visualization Analysis}.
\begin{figure}[t]
    \centering
    \includegraphics[width=0.475\textwidth]{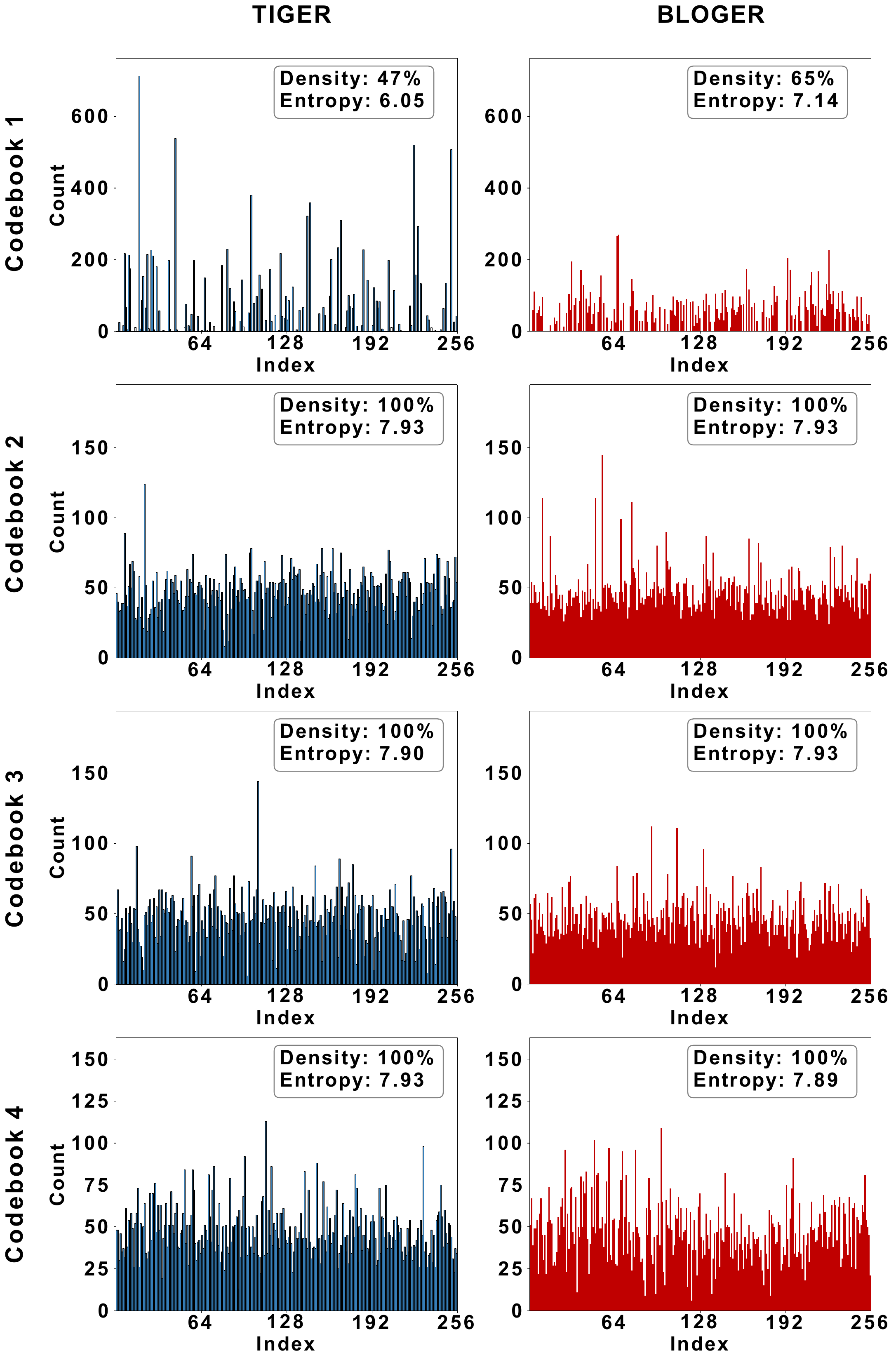}
    \caption{Codebook utilization comparison between TIGER and BLOGER, where “Codebook $i$” denotes the $i$-th codebook in the residual quantization process. “Density” is defined as the ratio of activated codewords to the total number of codewords, and “Entropy” measures the distribution of codeword usage, indicating how evenly the codewords are utilized.}
    \label{fig:vis}
    \Description{...}
\end{figure}
To gain deeper insights into the effectiveness of BLOGER, we perform a visualization analysis of the learned tokenizer. Specifically, we collect the token sequences of all items on Beauty and analyze the utilization patterns of the codebook at each level. The metrics we compute include: \textit{Density}, defined as the ratio of activated codewords to the total number of codewords; and \textit{Entropy}, which reflects the distribution of codeword usage, indicating how evenly the codewords are utilized. For comparison, we also include results for TIGER, which can be viewed as an ablated version of our framework with all specific designs removed.

The detailed results are presented in Figure~\ref{fig:vis}. As expected, the density of the first-level codebook is generally lower than that of the subsequent levels, and it often exhibits significant variation across different datasets. Notably, BLOGER demonstrates a substantial improvement over TIGER at the first level, whereas the higher-level codebooks show minimal differences. Specifically, in the first-level codebook, BLOGER increases Density from 47\% to 65\% and Entropy from 6.05 to 7.14. We attribute this to the first-level codebook carrying the most information due to residual operations, making it more responsive to additional supervision.

Considering that the primary distinction of BLOGER lies in the additional supervision of the tokenizer via the recommendation loss under a bi-level optimization framework, we can conclude that the recommendation loss acts like a \textbf{compass}, revealing hidden collaborative currents and correcting the tokenizer’s otherwise myopic, text-only view of the codebook landscape. By effectively incorporating this recommendation loss within a bi-level optimization framework, the tokenizer can better capture the underlying collaborative signals, thereby improving item representation quality and overall recommendation performance.

%% file: 6_related_work.tex
\section{Related Work}
In this section,  we navigate through existing research on generative recommendation, and bi-level optimization.

\subsection{Generative Recommendation.} 
Generative recommendation typically refers to a paradigm that enables direct prediction of target items without the need for exhaustive matching over the entire candidate set~\cite{HSTU,PinRec,GeneRec, OneRec,TIGER,lin2025igd}. 
Early studies explored this paradigm in LLM-based recommendation by incorporating textual descriptions~\cite{BIGRec, IDGenRec, LLM2BERT4Rec} or collaborative signals~\cite{LLM-RecSys-ID, P5,BinLLM,CoLLM} into the prompt to index items. For example, BIGRec~\cite{BIGRec} represents items using their titles and fine-tunes LLMs for recommendation under a bi-step grounding paradigm. IDGenRec~\cite{IDGenRec}, represents each item as a unique, concise, semantically rich, and platform-agnostic textual identifier composed of human-readable tokens, enabling better adaptation to LLM-based recommenders. 
BinLLM~\cite{BinLLM} transforms collaborative embeddings obtained from external models into binary sequences, thereby aligning item representations with formats readily consumable by LLMs. 

Later works explored formulating recommendation as a sequence-to-sequence task by integrating item tokenization with autoregressive generation. They typically treat the two components separately, which can be broadly categorized into two groups based on the optimization strategy. The first employs sequential optimization, where the tokenizer is first trained independently to produce static item identifiers, which are then used by the recommender for generation tasks~\cite{TIGER,LETTER,SEATER, EAGER,TokenRec,ColaRec,LCRec}. For example, TIGER~\cite{TIGER} represents each item with a Semantic ID—a tuple of codewords—and uses a Transformer to predict the next one. LETTER~\cite{LETTER} further introduces collaborative and diversity regularization to enhance the quality of the learned identifiers. 
\bym{QARM~\cite{QARM} mines a group of high-quality down-streaming task item-item pairs to guide the quantitative code learning.}
The second adopts alternating optimization, where the tokenizer and recommender are updated iteratively~\cite{ETEGRec}. For instance, ETEGRec~\cite{ETEGRec} incorporates two auxiliary losses to align the intermediate representations of the tokenizer and the recommender, facilitating end-to-end training through alternating updates of both components. 
In contrast, our proposed model BLOGER adopts a bi-level optimization framework, which more directly captures the dependency between item tokenization and recommendation generation.

\subsection{Bi-Level Optimization in Recommendation.}
Bi-level optimization is a class of problems involving two nested levels—an outer-level and a inner-level—where the objective and variables of the outer-level problem depend on the optimal solution of the inner-level~\cite{BLO,BLO2}. Owing to this hierarchical structure, bi-level optimization is naturally suited for modeling tasks with intrinsic interdependencies, and has been widely adopted in recommendations~\cite{LabelCraft,EBLO,BiLLP,CFGAN,Adv-MultVAE,APR}. For example, BOD~\cite{EBLO} utilizes bi-level optimization to adaptively learn sample weights for mitigating noise in implicit feedback. LabelCraft~\cite{LabelCraft} applies it to automatically generate reliable labels from raw feedback in short video recommendation, thereby achieving direct alignment with platform objectives. In contrast, we are the first to apply bi-level optimization to reshape generative recommendation, thereby opening up new possibilities for direct collaborative alignment.

Solving bi-level optimization problems is inherently challenging due to their nested structure and the implicit dependency of the outer-level objective on the inner-level solution~\cite{BLO2}. To tackle this, various gradient-based approaches have been proposed, all centered on efficiently computing the meta-gradient of the outer-level objective with respect to the inner-level optimum. These include explicit gradient updates~\cite{MAML} that explicitly approximate the inner-level solution by one-step optimizer updates, implicit function updates~\cite{IMAML} that exploit optimality conditions to avoid costly unrolling of inner-level iterations, and closed-form techniques~\cite{CloseForm} that yield exact gradients for specific problem classes. The tailored learning scheme within the BLOGER framework belongs to the first category and innovatively integrates gradient surgery techniques, enabling an efficient solution to our bi-level optimization problem.

%% file: 7_conclusion.tex
\section{Conclusion}
In this work, we reformulated generative recommendation as a bi-level optimization problem to explicitly capture the intricate interdependence between the tokenizer and the recommender. We proposed BLOGER, an \textit{efficient} and \textit{effective} framework that leveraged meta-learning and gradient surgery techniques to address this challenging optimization problem without introducing significant additional computational overhead, enabling direct alignment between tokenization and recommendation objectives and fostering mutual enhancement of both components. 
Interestingly, our experiments reveal that the performance gains primarily stem from improved codebook utilization. This suggests that the recommendation loss naturally acts like a compass, capturing collaborative signals and helping correct codebook allocation bias through gradient-based optimization.

For future work, we plan to further validate BLOGER’s generalization capabilities by conducting experiments on more diverse datasets, especially \textit{large-scale} industrial datasets, to better evaluate its scalability and practical applicability. Additionally, we will investigate practical deployment considerations to facilitate the adoption of BLOGER in production environments. Finally, we aim to develop tokenization strategies that are more suitable for large-scale streaming recommendation scenarios, and explore explicit or latent recommendation reasoning capabilities~\cite{latentr3,onerec-think}.